\pdfoutput=1
\documentclass[12pt,a4paper]{article}
\usepackage[T1]{fontenc}
\usepackage{graphicx}
\usepackage[compress]{scicite}
\usepackage{times}

\newenvironment{sciabstract}{\begin{quote}\bf}{\end{quote}}

\newcounter{lastnote}
\newenvironment{scilastnote}{%
\setcounter{lastnote}{\value{enumiv}}%
\addtocounter{lastnote}{+1}%
\begin{list}%
{\arabic{lastnote}.}
{\setlength{\leftmargin}{.22in}}
{\setlength{\labelsep}{.5em}}}
{\end{list}}

\title{Specific chemical reactivities of spatially separated 3-aminophenol conformers with cold Ca$^+$ ions}
\author{Yuan-Pin Chang,$\!^{1\dagger}$ Karol D\l ugo\l \k{e}cki,$\!^{1}$ Jochen K\"upper$^{1,2,3\,\ast}$,\\
   Daniel R\"osch,$\!^{4\dagger}$ Dieter Wild,$\!^{4}$ Stefan Willitsch$^{4\,\ast}$ \\[1ex]
   $^{1}$Center for Free-Electron Laser Science, DESY, \\ Notkestrasse 85, 22607 Hamburg, Germany \\[0.5ex]
   $^{2}$The Hamburg Center for Ultrafast Imaging, University of Hamburg, \\ Luruper Chaussee 149, 22761 Hamburg, Germany \\[0.5ex]
   $^{3}$Department of Physics, University of Hamburg, \\ Luruper Chaussee 149, 22761 Hamburg, Germany \\[0.5ex]
   $^{4}$Department of Chemistry, University of Basel, \\ Klingelbergstrasse 80, 4056 Basel, Switzerland \\[1em]
   $^\dagger$These authors contributed equally to this work. \\[0.5ex]
   $^\ast$To whom correspondence should be addressed; e-mail: \\ jochen.kuepper@cfel.de, stefan.willitsch@unibas.ch}%
\date{}%

\begin{document}
\baselineskip24pt
\maketitle
\begin{sciabstract}
   Many molecules exhibit multiple rotational isomers (conformers) that interconvert thermally and
   are difficult to isolate. Consequently, a precise characterization of their role in chemical
   reactions has proven challenging. We have probed the reactivity of specific conformers using an
   experimental technique based on their spatial separation in a molecular beam by electrostatic
   deflection. The separated conformers react with a target of Coulomb-crystallized ions in a trap.
   In the reaction of Ca$^+$ with 3-aminophenol, we find a twofold larger rate constant for the
   \textit{cis}- compared to the \textit{trans}-conformer (differentiated by the O-H bond
   orientation). This result is explained by conformer-specific differences in the long-range
   ion-molecule interaction potentials. Our approach demonstrates the possibility of controlling
   reactivity through selection of conformational states.
\end{sciabstract}

\noindent%
Recent progress in the cooling, manipulation and control of isolated molecules in the gas
phase~\cite{vandemeerakker12a,quemener12a,filsinger11a,willitsch12a} has paved the way for the study
of chemical processes at high levels of sensitivity, selectivity, and detail. Methods for the
slowing and merging of supersonic molecular beams have enabled precise characterizations of the role
of the collision energy and of the molecular quantum state in scattering and reactive
processes~\cite{gilijamse06a,kirste12a,henson12a}. Recent experiments with trapped, translationally
cold molecules and ions have provided insights into the quantum dynamics of chemical
reactions~\cite{ospelkaus10b} and the subtleties of intermolecular interactions~\cite{hall12a}.
These experiments have thus far been restricted to reactions involving atoms or small molecules with
simple geometric and quantum structures. The vast majority of molecules, however, possess a plethora
of internal degrees of freedom that are challenging to probe independently. In particular,
polyatomic molecules usually exhibit many rotational isomers (conformers) that interconvert with low
thermal barriers through rotations about covalent bonds. Recent years have seen impressive progress
in the spectroscopic characterization of specific
conformations~\cite{simons05a,Vries:ARPC58:585,rizzo09a}, their photoinduced
isomerization~\cite{dian04a,dian08a}, and the characterization of conformer-specific
photodissociation dynamics~\cite{park02a,kim07a}. However, studies of specific conformational
effects in bimolecular reactions are still sparse~\cite{khriachtchev09a,taatjes13a}, in particular
in the gas phase where the highest degree of control and therefore insight into fundamental reaction
mechanisms can be gained.

Here, we introduce a distinct approach for the study of conformational effects and
con\-for\-ma\-tion-dependent reactivities, i.\,e., rate constants, in bimolecular reactions. Our
method exploits the spatial separation of different conformers using inhomogeneous electric
fields~\cite{Filsinger:PRL100:133003,filsinger09a}. Internally cold molecules in a beam are
dispersed by an electrostatic deflector and are directed towards the reaction volume. There, they
interact with laser cooled ions in a Coulomb crystal, i.\,e., an ordered structure of
translationally cold ions in a trap~\cite{willitsch12a}. The Coulomb crystal constitutes a tightly
localized, stationary reaction target much smaller in extension than the conformationally separated
regions of the beam so that chemical reactions can be studied selectively with isolated conformers.

We studied the reaction of individual conformers of 3-aminophenol (AP) with Coulomb-crystallized
Ca$^+$ ions to probe conformation-dependent reactivities in a prototypical reaction between an
organic molecule and a metal ion in the gas phase. Reactions of this type are of interest, e.\,g.,
in the context of bond activation in catalysis \cite{Eller:1991gj,schwarz11a} and elucidating
interactions between metal ions and biomolecular building blocks \cite{Ryzhov:1999fs}. Fig.~1 shows
a schematic of the experimental setup. The vapor above a sample of AP heated to 145~$^{\circ}$C was
entrained in a pulsed supersonic expansion of neon at a stagnation pressure of 34 bar. AP exhibits
two distinct molecular conformations \textit{cis} and \textit{trans} that differ in the relative
orientation of the O-H bond with respect to the NH$_2$-group. The molecular geometries are depicted
in Fig.~2~A. The barrier to interconversion by rotation of OH about the C-O bond is
$\approx1500$~cm$^{-1}$~\cite{Robinson:JPCA108:4420}. Both conformers have significantly different
electric dipole moments of $0.77$~D and $2.33$~D for the \textit{trans} and \textit{cis}-species,
respectively, giving rise to their distinct Stark interactions with an electric
field~\cite{Filsinger:PCCP10:666}.

In the expansion, the AP molecules were adiabatically cooled to a population ratio of the
\textit{cis} and \textit{trans} conformers in the molecular beam of approximately $1:4$ and a
rotational temperature of $T_{\mathrm{rot}}=1.1$~K. Thus, over $99\,\%$ of the population of both
conformers was confined to the lowest rotational levels (rotational quantum numbers
$j_{\mathrm{AP}}{<\atop\sim}8$) and practically all molecules were in the vibrational and electronic
ground state. In the cold and collisionless environment of the molecular beam, interconversion
between the conformers did not occur. After passing two skimmers, the molecular beam entered an
electrostatic deflector consisting of a pair of 15~cm long electrodes to which potential differences
in the range of 5 to 13~kV were applied~\cite{Filsinger:JCP131:064309}. The shape of the electrodes
was designed to generate a strong inhomogeneous electric field with a nearly constant gradient along
the $y$ axis as depicted in Fig.~1. For both conformers, the Stark energies of all populated
rotational levels decrease with increasing field strength. The molecules are, therefore, deflected
towards regions of high electric field in the deflector~\cite{filsinger09a}. Because the electric
dipole moment is considerably larger for \textit{cis}-AP than for
\textit{trans}-AP~\cite{Filsinger:PCCP10:666}, the \textit{cis} species is more strongly deflected
than the \textit{trans} species resulting in a spatial separation of the two conformers in the
beam~\cite{filsinger09a}.

Fig.~2~A shows density profiles $n_{cis}(y)$ and $n_{trans}(y)$ for the deflected \textit{cis}- and
\textit{trans}-AP molecules, respectively. The profiles were obtained by measuring the ion signal
produced by conformer-selective resonance-enhanced multiphoton ionization
(REMPI)~\cite{filsinger09a} as a function of the vertical deflection coordinate when applying a
potential difference of $V_{\mathrm{defl}}=7.5$~kV to the deflector electrodes. The molecular beam
assembly was tilted incrementally in the vertical direction in order to scan the deflected molecular
beam over the position of the fixed ionization laser spot $82$~cm downstream from the exit of the
deflector. The deflection coordinate $y$ in Fig.~2 is defined as the vertical displacement of the
center of the nominally undeflected beam from the interaction point. The dashed lines represent the
results of Monte-Carlo trajectory simulations. In the calculations, the rotational temperature of
the molecules and one global scaling factor were adjusted to optimally reproduce the experimental
data. At low deflection coordinates the \textit{trans}-conformation dominates, whereas at the
highest deflection coordinates only the \textit{cis}-conformer is present.

For the reaction experiments, the REMPI spectrometer was replaced by a linear quadrupole ion trap
for the generation of Coulomb crystals of laser-cooled Ca$^+$ ions~\cite{willitsch12a}, see Fig.~1.
The Coulomb crystals consisting of typically 700 ions were imaged by a camera sampling the atomic
fluorescence generated by the laser cooling of the ions. The Ca$^+$ ions were exposed to AP
molecules from the deflected molecular beam. Because the AP molecules in the reaction volume were
replenished with each gas pulse, their number density was essentially constant over the measurement
time. Thus, pseudo-first-order reaction rate constants could be determined from the decrease of the
number of Ca$^+$ ions in the crystals~\cite{willitsch08a}. Product ions formed in the reaction
remained confined in the trap. They were sympathetically cooled by the interaction with the
remaining laser-cooled Ca$^+$ ions and localized at the edges of the Coulomb
crystals~\cite{willitsch12a}. These ions did not fluoresce and were only indirectly visible in the
images through a characteristic flattening of the Ca$^+$ crystal edges~\cite{willitsch08b} as shown
in Fig.~1.

The products of the reaction were analyzed using resonant-excitation mass spectrometry of the ions
in the trap, suggesting CaOH$^+$ or CaNH$_2^+$ ions and the corresponding 3-aminophenyl or
3-hydroxiphenyl radicals as the primary reaction products. In the experiment, the $(4s)\,^2S_{1/2}$,
$(4p)\,^2P_{1/2}$, and $(3d)\,^2D_{3/2}$ states of Ca$^+$ were populated as a consequence of the
laser cooling. The rate constant for reactions out of the excited $(4p)\,^2P_{1/2}$ state was found
to be two to three orders of magnitude larger than the rate coefficients for reactions out of the
two other states so that this process dominated the effective rates measured in the present study.
In the following, we focus on the Ca$^+(4p)$+AP reaction and all second-order rate constants quoted
below refer to this particular channel.

Fig.~2~B shows the total effective pseudo-first-order rate constant $k_{1,\mathrm{total}}$ as a
function of the deflection coordinate $y$. The measured rate constant profile
$k_{1,\mathrm{total}}(y)$ reflects both the density distribution of the conformers in the deflected
molecular beam as well as the conformer-specific second-order rate constants $k_{2,{cis}}$ and
$k_{2,{trans}}$ for the reactions of \textit{cis}-AP and \textit{trans}-AP, respectively, with
Ca$^+(4p)$:%
\begin{equation}
   k_{1,\mathrm{total}}(y)=k_{2,{cis}}~p_{4p}~n_{cis}(y)+k_{2,{trans}}~p_{4p}~n_{trans}(y).
   \label{eqn:rates}
\end{equation}
$p_{4p}$ is the population in the Ca$^+(4p)$ state. $k_{2,{cis}}$ and $k_{2,{trans}}$ were
determined from a global fit of Eq. 1 to reaction-rate profiles obtained at deflector voltages
$V_{\mathrm{defl}}=5$, $7.5$, $10$, and $13$~kV. The fit yielded the rate constants
$k_{2,{cis}}=(3.2\pm1.3)\times10^{-9}~\mathrm{cm^3~s^{-1}}$,
$k_{2,{trans}}=(1.5\pm0.6)\times10^{-9}~\mathrm{cm^3~s^{-1}}$ and the ratio
$k_{2,{cis}}/k_{2,{trans}}=2.1\pm0.5$ within a 95\,\% confidence interval. Fig.~2~C shows the total
second-order rate constant $k_{2,\mathrm{total}}(y)=x_{cis}(y)k_{2,{cis}}+x_{trans}(y)k_{2,{trans}}$
obtained from the fit, where $x_{cis}(y)$ and $x_{trans}(y)$ denote the mole fractions of
\textit{cis}-AP and \textit{trans}-AP in the molecular beam, respectively. This plot highlights the
change of the reaction rate as the beam composition evolves from \textit{trans} to \textit{cis}.

The rate constants $k_{2,{cis}}$ and $k_{2,{trans}}$ are of similar magnitude as the ones typically
obtained for ion-molecule reactions with capture-limited kinetics~\cite{rowe87a}. In these cases,
the reaction rates are dominated by long-range intermolecular interactions in the entrance channel
and are independent of the details of the short-range reaction mechanism. The rate constants can
then appropriately be modeled using adiabatic capture models~\cite{clary85a,stoecklin92a,troe87a}.
These models assume that the reaction happens with unit probability if the collision energy exceeds
the height of the centrifugal barrier in the entrance channel at a given value of the total angular
momentum. Fig.~3~A and B show the centrifugally corrected long-range potentials for different values
of the total angular momentum quantum number $J$ in the entrance channel of the reaction of Ca$^+$
with \textit{cis}- and \textit{trans}-AP, respectively. The potentials have been constructed from
the charge-dipole and the charge-induced dipole interactions, which represent the dominant
long-range forces relevant for the collisions~\cite{stoecklin92a}.

Due to the larger dipole moment of the \textit{cis}-conformer, the interaction potential is more
strongly attractive for \textit{cis}-AP than for \textit{trans}-AP. This can directly be seen from a
comparison of the $J=0$ potentials of the \textit{cis}- and \textit{trans}-conformers in Fig.~3~A
and B, respectively. As a result, the centrifugal barrier is more strongly suppressed in the
reaction with \textit{cis}-AP so that reactive collisions can occur up to higher maximum values
$J_\mathrm{max}$ of the total angular momentum quantum number $J$. At the collision energy of the
experiment, we find $J_\mathrm{max}=417$ and $J_\mathrm{max}=342$ for the \textit{cis}- and
\textit{trans}-conformers, respectively. Hence, in a classical picture the \textit{cis}-conformer
exhibits a larger maximum impact parameter $b_\mathrm{max}=J_\mathrm{max}/\mu{}v$ for the reaction
(where $\mu$ is the reduced mass and $v$ the collision velocity) and, therefore, an increased
reaction cross section $\sigma=\pi{}b_\mathrm{max}^2$. Averaging over all populated rotational
states, the adiabatic capture model predicts the rate constants
$k_{2,{cis}}=2.7\times10^{-9}~\mathrm{cm^3~s^{-1}}$ and
$k_{2,{trans}}=1.8\times10^{-9}~\mathrm{cm^3~s^{-1}}$. The ratio of the theoretical rate constants
is calculated to be $k_{2,{cis}}/k_{2,{trans}}=1.5$ in agreement with the experimental results.
Moreover, the absolute values of these capture rate constants are in very good agreement with the
experimentally obtained values. Thus, in the present case the increased reaction rate for the
\textit{cis} species can be traced back to its increased collision rate with Ca$^+$. This effect
results from conformation-dependent electrostatic properties of the molecules and the resulting
differences in the long-range interaction potentials. Short-range conformational effects play only a
minor role.

The isolation of individual conformers opens up new avenues to control the outcome of chemical
reactions by selecting conformer-specific pathways that result in different product distributions.
This capability has already been demonstrated in previous conformationally resolved
photodissociation experiments \cite{park02a, kim07a} and is now within reach for gas-phase reaction
experiments.

The present advances have become possible through the combination of electrostatic conformer
selection with highly sensitive Coulomb-crystal methods. We expect that the current methodology will
benefit fundamental reaction-dynamics studies as well as the investigation of a wide range of
ion-molecule reactions with relevance for catalysis and interstellar chemistry. Electrostatic
conformer selection is a widely applicable technique whenever the conformers present in the
molecular beam possess sufficiently different dipole moments. Even more advanced electric field
manipulation techniques for the separation of individual chemical
species~\cite{Filsinger:PRL100:133003, Trippel:PRA86:033202} or individual quantum
states~\cite{vandemeerakker12a, Nielsen:PCCP13:18971} have been demonstrated. In addition,
sympathetic cooling of ions is a near-universal technique that allows the generation of Coulomb
crystals of a wide range of atomic and molecular species with simultaneous preparation of their
internal quantum state~\cite{willitsch12a, tong10a}.

{\footnotesize%
   \bibliography{ca3ap}
   \bibliographystyle{Science}
}
\begin{footnotesize}
\begin{scilastnote}
\item[] \noindent {\bf Acknowledgements:} This work has been supported by the excellence cluster
   ``The Hamburg Center for Ultrafast Imaging -- Structure, Dynamics and Control of Matter at the
   Atomic Scale'' of the Deutsche Forschungsgemeinschaft, the Swiss National Science Foundation
   grant.\ nr.\ PP00P2\_140834, and the University of Basel. \\
\end{scilastnote}
\end{footnotesize}

\begin{figure}
   \centering
   \includegraphics[width=\linewidth]{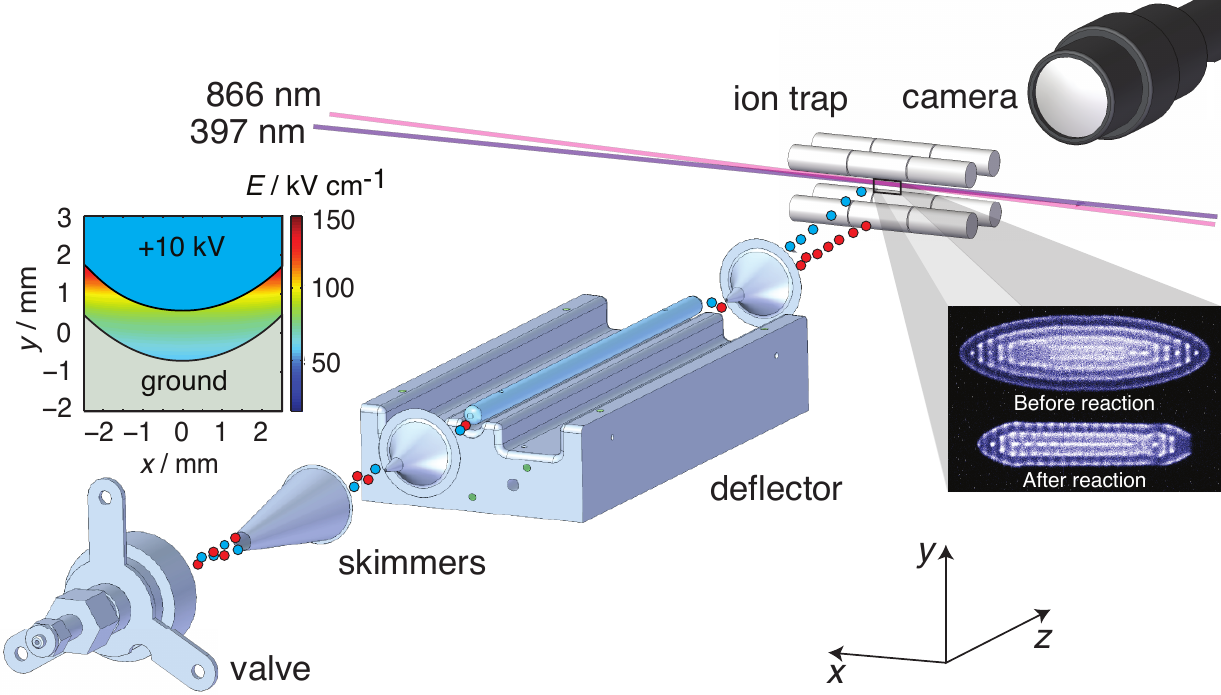}
   \caption{Experimental setup for studying conformer-selected chemical reactions. \textit{cis}-
      (blue spheres) and \textit{trans}- (red spheres) 3-aminophenol (AP) molecules are entrained in
      a molecular beam. The two conformers are deflected to different extents in the inhomogeneous
      electric field $E$ (left inset) of an electrostatic deflector. As a consequence, the molecular
      beam is split into different conformational components. The individual conformers are
      subsequently aimed at a stationary reaction target comprising a Coulomb crystal of
      laser-cooled Ca$^+$ ions in a trap (right inset). Conformer-specific rate constants are
      determined by monitoring the removal of Ca$^+$ ions from the Coulomb crystal as a function of
      the reaction time. 866~nm and 397~nm refer to the wavelengths of the Ca$^+$ cooling laser
      beams.}
   \label{fig:1}
\end{figure}

\begin{figure}
   \centering
   \includegraphics[width=\linewidth]{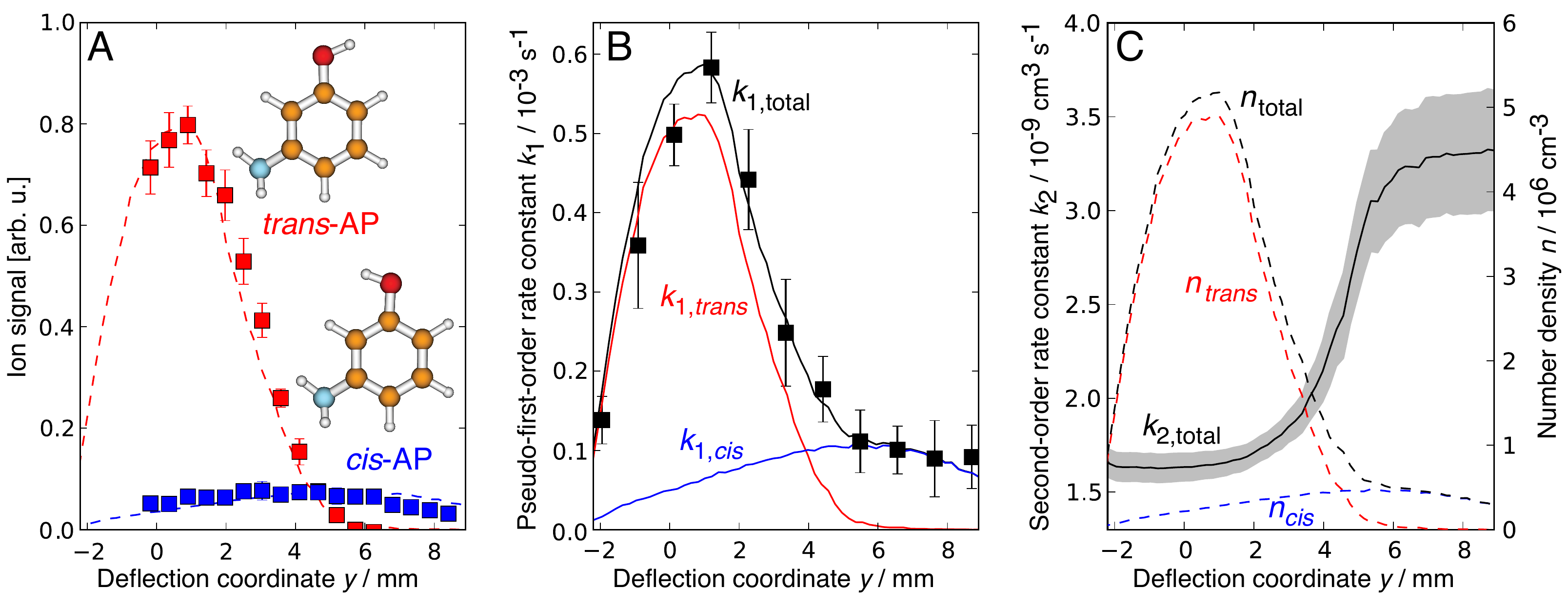}
   \caption{Reaction kinetics. (A) Density profiles of a deflected beam of \textit{cis}- and
      \textit{ trans}-AP (deflector voltage $V_\mathrm{defl}=7.5$~kV) measured by conformer-specific
      multi-photon ionization as a function of the molecular-beam deflection coordinate $y$.
      Squares: experimental data, dashed lines: Monte-Carlo trajectory simulations. (B)
      Conformer-specific (red and blue) and total (black) pseudo-first-order rate constants $k_1$
      for the reaction Ca$^+$ + \textit{cis}-/\textit{trans}-AP as a function of the beam deflection
      coordinate $y$ at $V_\mathrm{defl}=7.5$~kV. (C) Total bimolecular rate constant
      $k_\mathrm{2,total}(y)$ (black solid line) illustrating the increase in chemical reactivity as
      the predominant beam component changes from the \textit{trans} to the \textit{cis} conformer.
      Dashed lines: number densities of the two conformers. The data points represent the
      statistical averages of 1000 laser shots in (A) and a minimum of four reaction measurements in
      (B). Error bars indicate the corresponding 95\,\% confidence intervals. The gray area in (C)
      represents the 95\,\% confidence interval of $k_\mathrm{2,total}$.}
   \label{fig:2}
\end{figure}

\begin{figure}
   \centering
   \includegraphics[width=\linewidth]{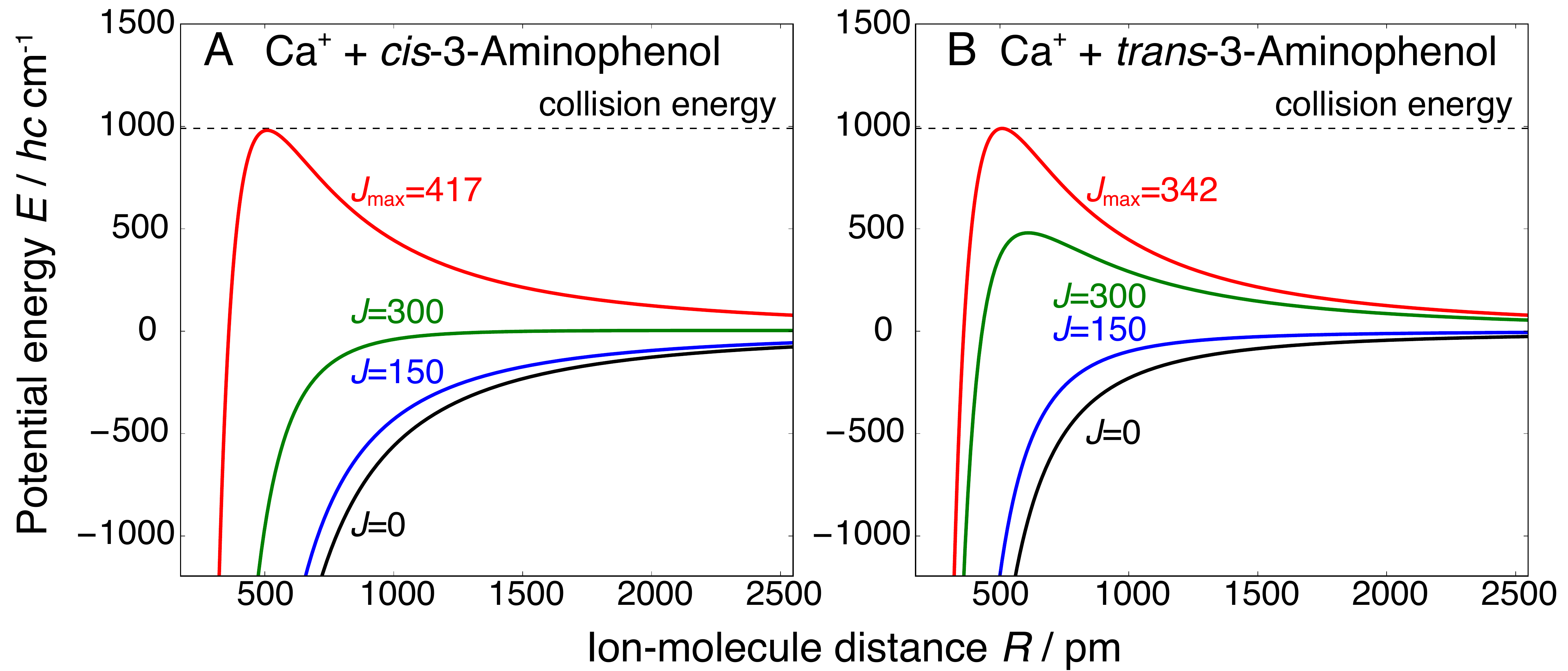}
   \caption{Centrifugally corrected potential energy curves for rotationless (A) \textit{cis}- and
      (B) \textit{trans}-AP molecules reacting with Ca$^+$.}
   \label{fig:3}
\end{figure}

\end{document}